\documentclass[twocolumn, longauthor]{aastex631}

\usepackage{xcolor}
\usepackage{comment}
\usepackage{graphicx}
\usepackage{dcolumn}
\usepackage{bm}
\usepackage{hyperref}
\DeclareUnicodeCharacter{2009}{\,}

\begin{document}

\title{Librating Kozai-Lidov Cycles with a Precessing Quadrupole Potential are Analytically Approximately Solved}

\correspondingauthor{Ygal Y. Klein}
\email{ygalklein@gmail.com}

\author[0009-0004-1914-5821]{Ygal Y. Klein}
\affiliation{Dept. of Particle Phys. \& Astrophys., Weizmann Institute of Science,
 Rehovot 76100, Israel}

\author[0000-0003-0584-2920
]{Boaz Katz}
\affiliation{Dept. of Particle Phys. \& Astrophys., Weizmann Institute of Science,
 Rehovot 76100, Israel}







\begin{abstract}

 The very long-term evolution of the hierarchical restricted three-body problem with a slightly aligned precessing quadrupole potential is investigated analytically for librating Kozai-Lidov cycles (KLCs). \citet{klein2023} presented an analytic solution for the approximate dynamics on a very long timescale developed in the neighborhood of the KLCs fixed point where the eccentricity vector is close to unity and aligned (or anti aligned) with the quadrupole axis and for a precession rate equal to the angular frequency of the secular Kozai-Lidov Equations around this fixed point. In this Letter, we generalize the analytic solution to encompass a wider range of precession rates. We show that the analytic solution approximately describes the quantitative dynamics for systems with librating KLCs for a wide range of initial conditions, including values that are far from the fixed point which is somewhat unexpected. In particular, using the analytic solution we map the strikingly rich structures that arise for precession rates similar to the Kozai-Lidov timescale (ratio of a few).

\end{abstract}



\section{Introduction} \label{sec:intro}

In this Letter we generalize and study the applicability of the analytical solution introduced by \citet{klein2023} for the long-term evolution of a test particle in a Keplerian orbit perturbed by a slightly aligned precessing quadrupole potential. The system under consideration involves a test particle orbiting a central mass $M$ on a Keplerian orbit with semimajor axis $a$ that is perturbed by an external quadrupole potential given by 
\begin{equation}
\Phi_{\text{outer}}=\frac{\Phi_0}{a^2}\left[3\left(\mathbf{\hat{j}}_{\text{outer}}\cdot\mathbf{r}\right)^{2}-r^{2}\right]\label{eq:Potential_phi}
\end{equation}
where $\Phi_0$ is constant and $\mathbf{\hat{j}}_{\text{outer}}$ is a time dependent unit vector which precesses around the \textit{z}-axis at a constant rate $\beta$ with a constant inclination $\alpha$:
\begin{equation}
 \mathbf{\hat{j}}_{\text{outer}}=\left(\begin{array}{c}
   \sin\alpha\cos\left(\beta\tau\right) \\
   -\sin\alpha\sin\left(\beta\tau\right) \\   
   \cos\alpha
  \end{array}\right)\label{eq:jOuter_as_a_function_of_tau}
\end{equation}
where $\tau\equiv\frac{t}{t_{\text{sec}}}$ and $t_{\text{sec}}=\frac{\sqrt{GMa}}{\Phi_0}$ is the secular timescale.

For $\beta=0$ the potential remains constant over time and the system represents the scenario of a hierarchical three-body system, where the gravitational potential has been expanded to quadrupole order in terms of the separations ratio and averaged over the outer period. This problem has been analytically solved using the secular approximation (\citet{kozai62,lidov1962}, for a recent review see \citet{naoz2016}) resulting with the periodic Kozai-Lidov cycles (KLCs) which reach high eccentricities for high initial inclinations\footnote{ See recent historical overview including earlier relevant work by \citet{vonZeipel1910} in \citet{ito_2019}.}.

The emergence of high eccentricity for the inner binary of a hierarchical system in a KLC has been proposed as a crucial factor in the possible formation channel of a wide variety of astrophysical phenomena such as hot/warm planets \citep{naoz2011,katz2011,fabrycky2007}, planets around white dwarfs \citep{munoz20,oconnor21,stephan21}, Type Ia supernovae through the merger or collision of white dwarfs in multiple systems \citep{thompson2011,katz2012}, gravitational wave emission through the merger of black holes or neutron stars \citep{antonini2012,liu2018} and the formation of close binaries \citep{antonini2012}.

Previous numerical investigations have demonstrated that in systems with higher multiplicities, whether they include an additional perturber, account for the effect of the host galaxy, or consider the asphericity of star clusters, the phase space that allows for high eccentricities is more extensive when compared to equivalent triple systems \citep{pejcha2013,hamers2015,hamers2017a,petrovich2017,fang2018,grishin18,liu2019,safarzadeh2020,hamers2020,bub2020,grishin22}. The simplest case for rich dynamics in a higher multiplicity system is a case where the inner binary's angular momentum is negligible (the test particle assumption)
and there are no KLC dynamics for the other binaries in the systems. This simplest case of higher multiplicity (with some restricting assumptions) is equivalent to the precessing quadrupole studied in this Letter, Equations \ref{eq:Potential_phi} and \ref{eq:jOuter_as_a_function_of_tau}, and numerical explorations have validated that even for low alignments it exhibits a wide range of initial conditions leading to high eccentricity outcomes  \citep{hamers2017,petrovich2017}.

The dynamics of the test particle can be parameterized
by two dimensionless orthogonal vectors $\mathbf{j}=\mathbf{J}/\sqrt{GMa}$, where $\mathbf{J}$ is the specific angular momentum vector, and
$\mathbf{e}$ a vector pointing in the direction of the pericenter
with magnitude $e$. In the secular approximation, where the equations are averaged over the orbit, $a$ is constant with time while $\mathbf{j}$ and
$\mathbf{e}$ evolve according to
\begin{eqnarray}
\frac{d\mathbf{j}}{d\tau}=&\frac{3}{4}\left(\left(\mathbf{j}\cdot\mathbf{\hat{j}}_{\text{outer}}\right)\mathbf{j}-5\left(\mathbf{e}\cdot\mathbf{\hat{j}}_{\text{outer}}\right)\mathbf{e}\right)\times\mathbf{\hat{j}}_{\text{outer}} \cr
\frac{d\mathbf{e}}{d\tau}=&\frac{3}{2}\left(\mathbf{j}\times\mathbf{e}\right)-\frac{3}{4}\left(5\left(\mathbf{e}\cdot\mathbf{\hat{j}}_{\text{outer}}\right)\mathbf{j}-\left(\mathbf{j}\cdot\mathbf{\hat{j}}_{\text{outer}}\right)\mathbf{e}\right)\times\mathbf{\hat{j}}_{\text{outer}}. \cr
 \label{eq:secular_equations}
\end{eqnarray}

As mentioned above, in the periodic analytically solved KLCs the external potential is constant in time implying an axisymmetric potential and admitting a constant of the motion, $\mathbf{j}\cdot\mathbf{\hat{j}}_{\text{outer}}$. In this case a second constant of motion can be obtained from the double averaged potential and $\mathbf{j}\cdot\mathbf{\hat{j}}_{\text{outer}}$ 
\begin{equation}
C_{K}=e^{2}-\frac{5}{2}\left(\mathbf{e}\cdot\mathbf{\hat{j}}_{\text{outer}}\right)^{2}. \label{eq:CK}
\end{equation}
The dynamics of a KLC are determined by the values of the constants $C_K$ and $\mathbf{j}\cdot\mathbf{\hat{j}}_{\text{outer}}$. When $C_K<0$, the argument of pericenter of the Keplerian orbit librates around $\frac{\pi}{2}$ or $-\frac{\pi}{2}$ (\textit{librating} cycles), and when $C_K>0$, it goes through all values $\left[0,2\pi\right]$ (\textit{rotating} cycles).

In the problem we study here with $\beta>0$, i.e $\mathbf{\hat{j}}_{\text{outer}}$ is precessing at a constant rate, $C_K$ and $\mathbf{j}\cdot\mathbf{\hat{j}}_{\text{outer}}$ are no longer constant. We note that in the presence of the third order term (octupole) in the expansion of the potential a change in $C_K$ and $\mathbf{j}\cdot\mathbf{\hat{j}}_{\text{outer}}$ leading to high eccentricities and orbit flips can be obtained \citep{naoz2011,katz2011,lithwick2011,teyssandier2013,li2014a,naoz2016,tan2020,lei2022,huang2022}. In the precessing potential studied here the change in $C_K$ and $\mathbf{j}\cdot\mathbf{\hat{j}}_{\text{outer}}$ is obtained at the quadrupole level \citep{petrovich2017}.

A global constant of motion of Equations \ref{eq:jOuter_as_a_function_of_tau}-\ref{eq:secular_equations} can be derived as follows (equivalent to the constant $\tilde{H}_{\text{rot}}$ presented in \cite{petrovich2017}). Time independent equations can be obtained by transforming to a reference frame rotating with the perturbing potential. This is equivalent to adding $\left(-\beta j_z\right)$ to the Hamiltonian \citep{tremaine2014} and implies the following constant of motion (equal to the resulting time independent Hamiltonian up to additive and multiplicative constants)
\begin{equation}
    C_K + \frac{1}{2}\left(\mathbf{j}\cdot\mathbf{\hat{j}}_{\text{outer}}\right)^{2} - \frac{4}{3}\beta j_z. \label{eq:global_constant}
\end{equation}

Since this expression is independent of the reference frame it represents a global constant of Equations \ref{eq:jOuter_as_a_function_of_tau}-\ref{eq:secular_equations} as can be readily checked directly\footnote{We thank Scott Tremaine for suggesting to consider a rotating frame.}. This constant applies for all initial conditions and any value of $\alpha$.

As outlined in \citet{klein2023}, when $0<\alpha\ll1$ the dynamics on short time scales adheres to the conventional KLC with two conserved quantities: $\mathbf{j}\cdot\mathbf{\hat{j}}_{\text{outer}} \approx j_{z}$ and $C_K$ (Equation \ref{eq:CK}). On longer time scales $j_{z}$ and $C_{K}$ evolve (satisfying the global constant of Equation \ref{eq:global_constant}). In this Letter, our emphasis is directed towards the circumstances conducive to high eccentricities, which correspond to low $\left|j_z\right|$ (extremely high eccentricities correspond to a zero crossing of $j_z$, e.g \citet{naoz2011}). A measure of the range of initial $j_z$ that can lead to a zero crossing outcome is the magnitude of variation observed in $j_z$ in the vicinity of $j_z=0$ \citep{katz2011,luo2016,haim2018}. This metric will serve as the primary quantitative parameter that we will examine, both numerically and analytically.

In \citet{klein2023} we presented an analytic solution to the KLC averaged equations of this problem for $\alpha\ll1$ in a restricted regime close to the KLC fixed point of $C_K=-1.5$, corresponding to  $\left(\mathbf{e}\cdot\mathbf{\hat{j}}_{\text{outer}}\right)^{2}=1$, $\left|i\right|=\frac{\pi}{2}$ and $j=0$ (i.e $e=1$). The effective averaged equations and analytic solution in \citet{klein2023} were derived for a specific value of precession rate
\begin{equation}
 \beta=\beta_{0}\equiv\sqrt{\frac{135}{16}}\approx2.9\label{eq:beta0},
\end{equation}
which is the natural resonant frequency of the equations of motion in the neighborhood of this fixed point ($\mathbf{e}$ close to unity and aligned or anti-aligned with $\mathbf{\hat{j}}_{\text{outer}}$ and $\mathbf{j} \approx 0$). The success of the solution to capture and approximately describe the long-term evolution of the dynamics was shown for examples with $\beta$ close to $\beta_0$ and initial conditions close to the fixed point.

In this Letter, we generalize the KLC averaged equations and their analytic solution for a wider range of $\beta$ values and compare it to the numerical results for a wide range of $\alpha$ and $\beta$. We explore a large phase space of initial conditions close and far from the fixed point $C_K=-1.5$ but we restrict our analysis to librating KLCs (i.e $C_K<0$ at all times), $\alpha\ll1$ and $\left|j_z\right|\ll1$.

\section{Numerical Results\label{sec:numerical setup}}

We numerically integrate Equations \ref{eq:jOuter_as_a_function_of_tau}-\ref{eq:secular_equations}  (up to $\tau=300$), with varying initial conditions for the vectors $\mathbf{e}$ and $\mathbf{j}$, as well as different values of $\alpha$ and $\beta$. Our analysis is restricted to simulations where the parameter $C_K$ remains negative throughout its temporal evolution, implying persistent libration in the KLC conditions. With an emphasis on the low $\left|j_z\right|$ region (favorable for inducing high eccentricity), we set the initial value of $j_z$ to zero. This implies that in all the results shown, the initial value of $C_K$, $C^0_K$, is equal to the global constant in Equation \ref{eq:global_constant} up to first order in $\alpha$. To cover the multidimensional phase space of initial conditions, for each combination of $\alpha$ and $\beta$ values under investigation, we generate 1000 instances of initial conditions. Every instance is produced by randomly selecting three values for $e_x$, $e_y$, and $j_x$ from a uniform distribution within the interval $\left[-1, 1\right]$. We then determine $e_z$ and $j_y$ by demanding the conditions $\mathbf{e} \cdot \mathbf{j} = 0$ and $e^2 + j^2 = 1$, selecting one solution at random from the two possible solutions. If no solution exists - a new random selection of $e_x$, $e_y$, and $j_x$ is made instead, implying 1000 instances of viable initial conditions. If $C^0_K>0$ or $C_K$ becomes positive during the evolution we discard the simulation from the analysis. The number of presented integrations is therefore less than $1000$ and is indicated explicitly for each result shown (top of the relevant subplots in the figures below).

\begin{figure}
 \begin{centering}
 \includegraphics[scale=0.21]{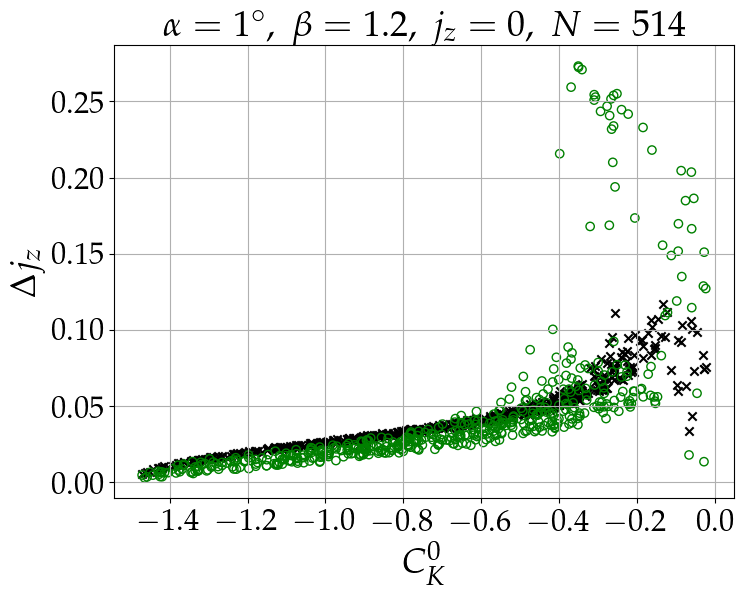}
  \includegraphics[scale=0.21]{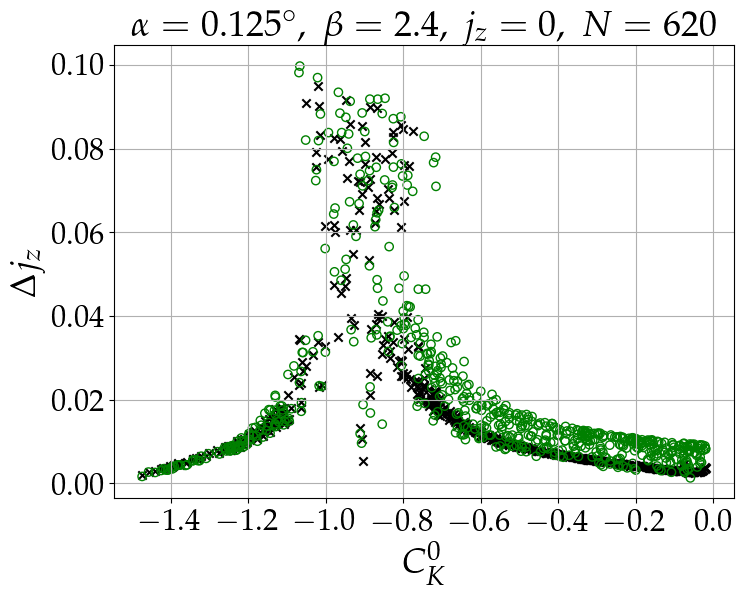}
  \par\end{centering}
 \begin{centering}
 \includegraphics[scale=0.21]{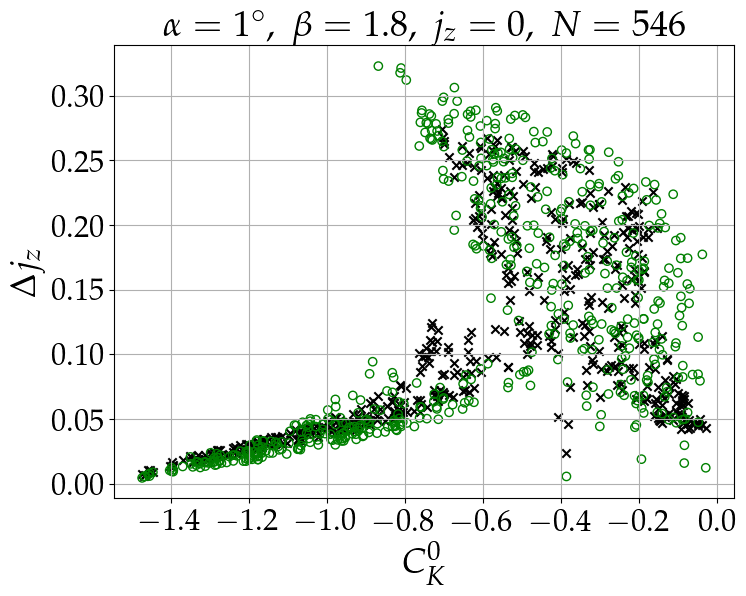}
  \includegraphics[scale=0.21]{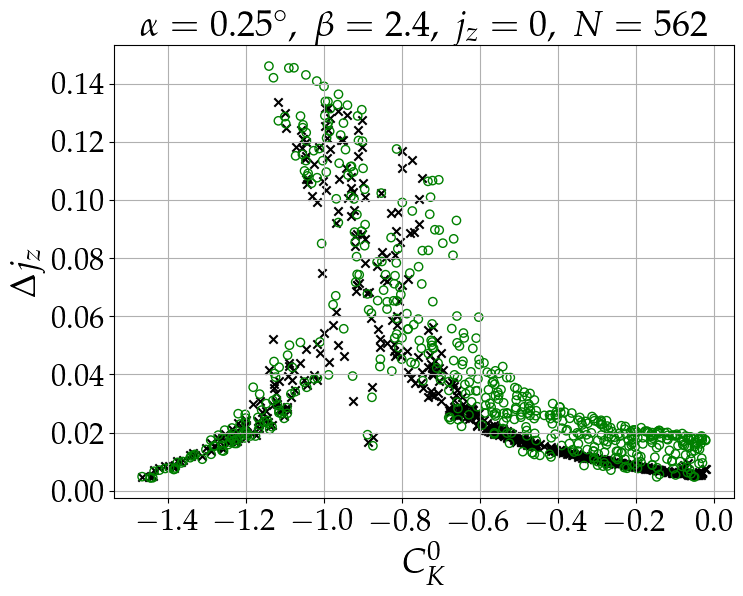}
  \par\end{centering}
   \begin{centering}
 \includegraphics[scale=0.21]{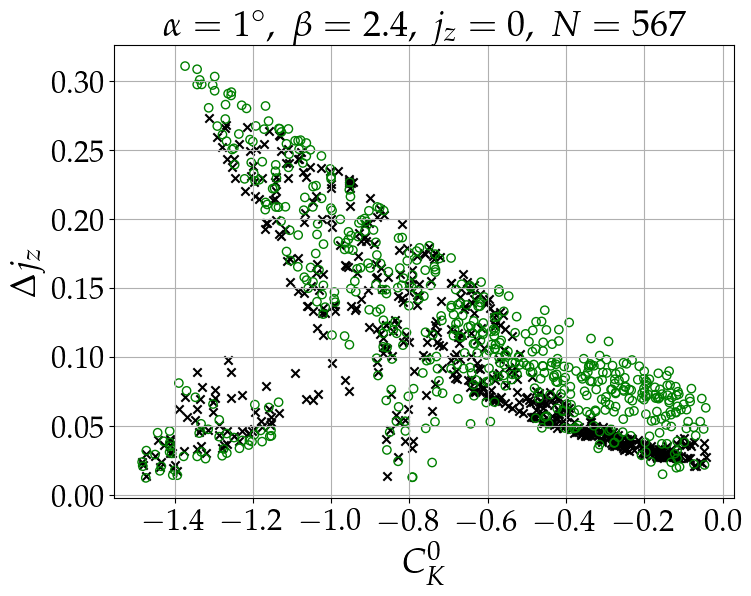}
 \includegraphics[scale=0.21]{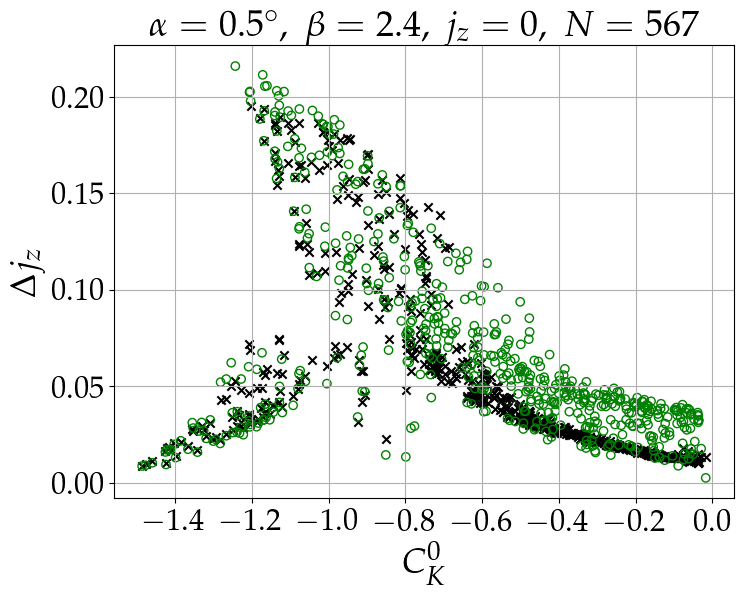}
  \par\end{centering}
   \begin{centering}
   \includegraphics[scale=0.21]{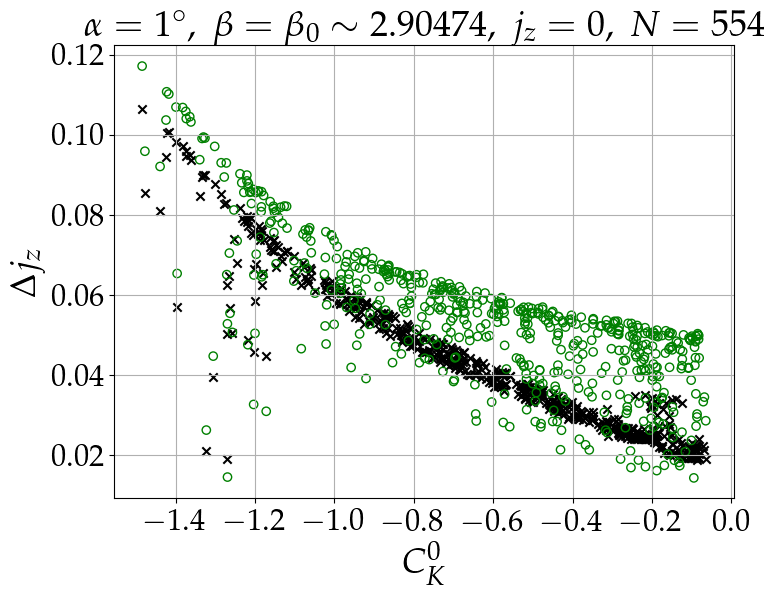}
   \includegraphics[scale=0.21]{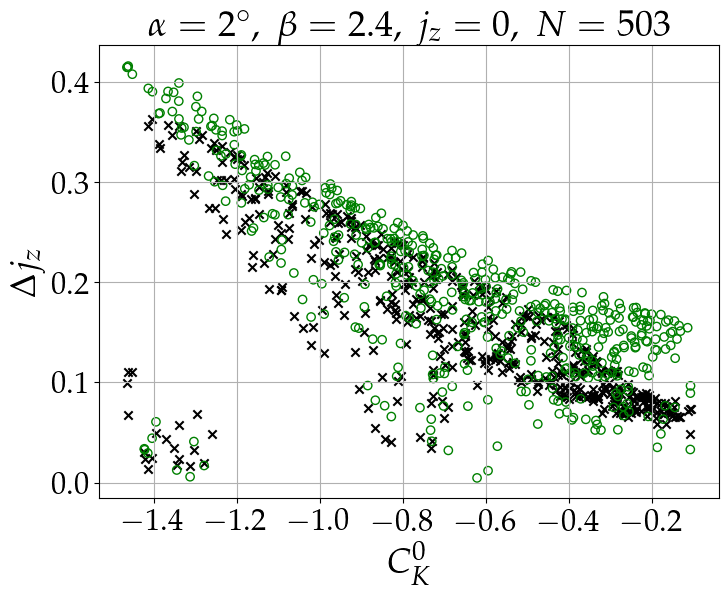}
  \par\end{centering}
     \begin{centering}
   \includegraphics[scale=0.21]{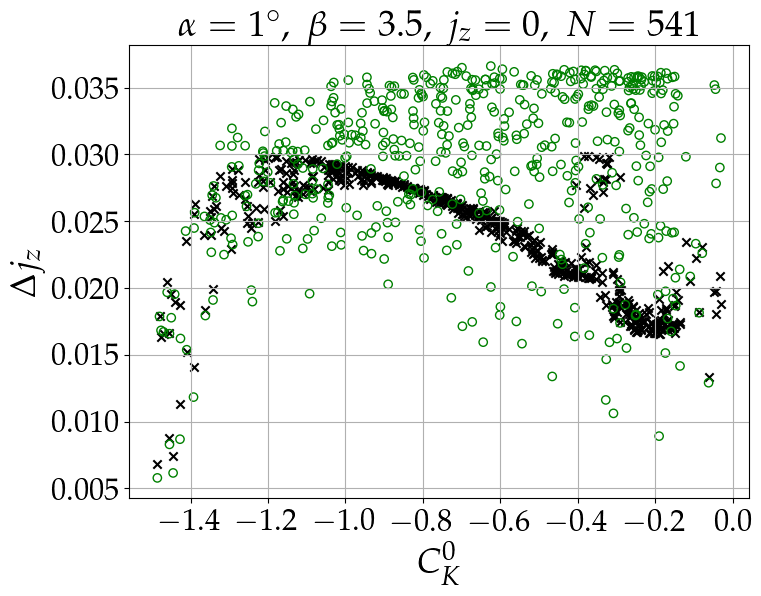}
   \includegraphics[scale=0.21]{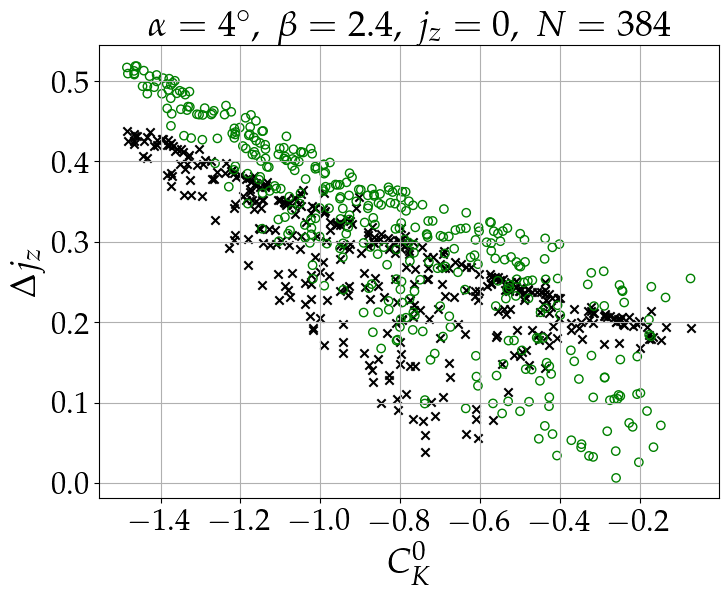}
  \par\end{centering} 
 \caption{Amplitude of the change in $j_z$ vs. initial $C_K$ (equal to the global constant in Equation \ref{eq:global_constant} with $j^0_z=0$ up to first order in $\alpha$) for librating KLCs. Left panels: constant $\alpha=1^{\circ}$ and different values of $\beta$ (written explicitly above each subplot). Right panels: constant $\beta=2.4$ and different values of $\alpha$ (logarithmically spaced). Results from a full numerical solution of Equations \ref{eq:jOuter_as_a_function_of_tau}-\ref{eq:secular_equations} are marked using black crosses. Green circles denote the prediction of the analytic solution using Equations \ref{eq:C4}, \ref{eq:energy} and \ref{eq:Potential_V} (see section \ref{sec:general beta equations and solution}). The number $N$ marked above each plot is the number of initial conditions shown. This number is the number of initial conditions that are librating KLCs among a random sample of $1000$ realizations of initial conidtions.}\label{fig:jz_amplitude_vs_CK_always_librating}
\end{figure}

The amplitude of change in $j_z$ throughout the numerical integration ($\Delta j_z=j^{\text{max}}_z-j^{\text{min}}_z$) versus $C^0_K$ is shown for various values of $\alpha$ and $\beta$ in Figure \ref{fig:jz_amplitude_vs_CK_always_librating} using black crosses. For each pair of $\alpha$ and $\beta$ values the results are shown in one subplot. In the left column of panels, we maintain $\alpha=1^{\circ}$ and explore different $\beta$ values. In the right column of panels, we set $\beta=2.4$ and examine different $\alpha$ values with equal logarithmic spacing.

Note the following striking features of the numerical results shown in Figure \ref{fig:jz_amplitude_vs_CK_always_librating}: (a) $j_z$ can change considerably (even) for low $\alpha$, an effect that is quickly suppressed when $\beta$ becomes larger than $\beta_0$. (b) $\Delta j_z$ is determined mainly by $C^0_K$. (c) The relationship between $\Delta j_z$ and $C^0_K$ shows intricate and non-uniform patterns: Various correlations are observed, distinct subdivisions into separate sub-categories are evident, notable levels of dispersion occur, there are instances of remarkably narrow correlation, and abrupt localized increments emerge.

The green circles in Figure \ref{fig:jz_amplitude_vs_CK_always_librating} denote results from the analytic solution outlined in the upcoming sections. Note the strong resemblance for most (but not all) of the distinctive characteristics between the two solutions. A detailed comparison is discussed in section \ref{sec:comparison}.

\section{Averaged Equations and Analytic Solution\label{sec:general beta equations and solution}} 

As demonstrated in \citet{klein2023}, close to the fixed point $C_K=-1.5$, up to first order in $\alpha$ and within the low $\left|j_z\right|$ regime, Equations \ref{eq:jOuter_as_a_function_of_tau}-\ref{eq:secular_equations} turn out to have a structure of a sinusoidal-driven harmonic oscillator equation for $e_x$ and $e_y$, under the assumption of a constant $e_z$. This simplification leads to the following equations: $\ddot{e}_{x}=\omega_{0}^{2}\left(L\cos\left(\omega\tau\right)-e_{x}\right)$ and $\ddot{e}_{y}=\omega_{0}^{2}\left(L\sin\left(\omega\tau\right)-e_{y}\right)$
where $\omega=\beta,L=e_{z}\alpha$ and $\omega_{0}=\beta_0\left|e_z\right|$ (see Equations 4-9 in \citet{klein2023}).

As a consequence, a measure of the proximity to resonance for these driven harmonic oscillators involves evaluating the difference $\omega^2_0-\omega^2$. In the context of a standard driven harmonic oscillator, both $\omega_0$ and $\omega$ remain constant over time, thereby establishing a fixed state of proximity to resonant behavior. However, in our examined scenario, $\omega=\beta$ remains unchanged while $\omega_0=\beta_0 \left|e_z\right|$ can slowly evolve, in addition to variations in the instantaneous $e_z$ induced by the KLC itself. In the following we shall average the equations over KLCs. The deviation from resonance, $\omega^2_0-\omega^2$, is represented by one of the slow dynamical variables that we follow (see Equation \ref{eq:delta} below). Non-trivial dynamics are obtained when the system approaches or even crosses resonance as the KLC-averaged value of $e_z$ gradually changes over time. 

\subsection{Ansatz (as in \citet{klein2023})\label{subsec:ansatz}}

For convenience we lay out in this subsection definitions and equations that were given in \citet{klein2023}: We use the following ansatz for the vector $\mathbf{e}$
in the $x-y$ plane motivated by the driven harmonic oscillator structure of the equations near the fixed point of $C_K=-1.5$ close to resonance. At any time $\tau$, the projection of the vector
$\mathbf{e}$ on the $x-y$ plane can be presented as a point moving
on a slowly evolving ellipse with semimajor axis $a$ inclined with an angle $\theta$ with respect to the \textit{x}-axis and semiminor axis $b$ centered at the origin, i.e,
\begin{equation}
 \mathbf{e}_{x-y}=\alpha^{\frac{1}{3}}\left(\begin{array}{cc}
   \cos\theta, & -\sin\theta \\
   \sin\theta, & \cos\theta
  \end{array}\right)\left(\begin{array}{c}
   a\cos\left(\hat{\beta}\hat{\tau}+\phi\right) \\
   b\sin\left(\hat{\beta}\hat{\tau}+\phi\right)
  \end{array}\right)\label{eq:exy-ansatz}
\end{equation}
where $\phi$ is a slowly dynamically evolving phase and
\begin{equation}
 \hat{\tau}=\frac{1}{2}\alpha^{\frac{2}{3}}\tau, \ \hat{\beta}=2\alpha^{-\frac{2}{3}}\beta.
 \label{eq:tau_hat_beta_hat}
\end{equation}
We use the following dynamical parameter
\begin{equation}
 \delta=\alpha^{-\frac{2}{3}}\frac{1}{\beta_{0}}\left(\left(\beta_{0}\bar{e}_{z}\right)^{2}-\beta^{2}\right)\label{eq:delta}
\end{equation}
measuring the proximity to resonance in the driven harmonic oscillator (with $\omega=\beta$ fixed but $\omega_0=\beta_0\left|\bar{e}_{z}\right|$ slowly evolving with time), 
where $\bar{e}_{z}$ is the averaged value of $e_{z}$ over KLC which satisfies
\begin{equation}
 e_{z}=\bar{e}_{z}+\frac{\alpha^{\frac{2}{3}}}{6}\left(a^{2}-b^{2}\right)\cos\left(2\left(\hat{\beta}\hat{\tau}+\phi\right)\right).\label{eq:instantanous_ez}
\end{equation}

\subsection{Averaged Equations \label{subsec:averaged equations general beta}}

In what follows the equations presented are new, with $\beta$ not restricted to equal $\beta_0$ but assumed to be in the vicinity of $\beta_0\left|e_z\right|$.

Using the following slow variables 
\begin{eqnarray}
 s=-\frac{\beta_0}{\beta}\text{sign}\left(e_z\right)\gamma\left(a-b\right)\sin\left(\theta-\phi\right)\\
 c=-\frac{\beta_0}{\beta}\text{sign}\left(e_z\right)\gamma\left(a-b\right)\cos\left(\theta-\phi\right)\label{eq:s_and_c}
\end{eqnarray}
where 
\begin{equation}
 \gamma=\beta^2_0+\frac{5}{3}\beta^2\label{eq:gamma},
\end{equation}
the following set of ODEs is obtained:
\begin{eqnarray}
 \frac{d\left(\theta+\phi\right)}{d\hat{\tau}} & =\frac{\beta_0}{\beta}\delta\label{eq:theta_minus_phi_dot} \\
 \frac{d\left(a+b\right)}{d\hat{\tau}}     & =0\label{eq:a_plus_b_dot},
\end{eqnarray}
and 
\begin{eqnarray}
 \dot{\delta}=&s\label{eq:deltaDot}\\
 \dot{s}=&-\frac{\beta_0}{\beta}\left(2\beta_{0}\gamma\frac{\beta_0}{\beta}+\delta c\right)\label{eq:splusDot}\\
 \dot{c}=&\frac{\beta_0}{\beta}\delta s\label{eq:cPlusDot}
\end{eqnarray}
where $\dot{}$ denotes a derivative with respect to $\hat{\tau}$.

Note that $a+b$ is a constant of motion and the equations for $\delta$, $s$ and $c$ form a closed subset.

Using the demand that $\mathbf{j}\cdot\mathbf{e}=0$ we deduce that the following combination of $j_z$ and $\delta$ is constant:
\begin{equation}
j_z+\sqrt{\frac{5}{3}}\frac{\beta}{\gamma}\alpha^{\frac{2}{3}}\delta=\text{const.},\label{eq:C4}
\end{equation}
enabling $j_z$ to be readily obtained at any time utilizing the initial conditions and the value of $\delta$.

\subsection{Analytic Solution\label{subsec:Analytic-Solution}}

The averaged equations, Equations \ref{eq:deltaDot}-\ref{eq:cPlusDot}, admit two constants of the motion, denoted $C_{1}$ and $C_{2}$,
\begin{eqnarray}
 C_{1}=&\frac{1}{2}\delta^{2}+\text{sign}\left(e_z\right)\gamma\left(a-b\right)\cos\left(\theta-\phi\right)\label{eq:C1}\\
 C_{2}=&\left(a-b\right)^{2}+4\beta_0\frac{\delta}{\gamma}\label{eq:C2}
\end{eqnarray}
implying that the evolution of the three variables $\delta,a-b$ and $\theta-\phi$ is periodic.

The resulting evolution of $\delta$ is equivalent to the dynamics of a particle moving in a one dimensional potential with a constant energy
\begin{equation}
 E=\frac{1}{2}\dot{\delta}^{2}+V=\frac{1}{2}\left(\frac{\beta_0}{\beta}\gamma\right)^{2}C_{2}\label{eq:energy}
\end{equation}
where
\begin{equation}
 V=\left(\frac{\beta_0}{\beta}\right)^{2}\left(2\beta_0\gamma\delta-\frac{1}{2}C_{1}\left(\delta^{2}-C_1\right)+\frac{1}{8}\delta^{4}\right)\label{eq:Potential_V}.
\end{equation}

The pair of constants, $C_1$ and $C_2$, which determine $V$ and $E$ offer a comprehensive analytical solution for the evolution of $\delta$. Consequently, this solution allows for the derivation of the evolution of all the slowly evolving variables. Note that the energy and potential are defined slightly different than in \citet{klein2023} (Equations 29-30 there) with an extra added constant $\frac{1}{2}\left(\frac{\beta_0}{\beta}C_1\right)^2$  so $E$ would depend on $C_2$ only. This addition of a constant value does not alter the results which are presented as a function of $C_1$ and $C_2$ and could have been added there.

Equations \ref{eq:energy} and \ref{eq:Potential_V} determine the amplitude of change in $\delta$ analytically. Using Equation \ref{eq:C4} we obtain the analytic prediction for $\Delta j_z$ which is plotted using green circles in Figure \ref{fig:jz_amplitude_vs_CK_always_librating}.

We note that at first glance, the analytic solution remarkably agrees with the general trends and most of the striking features and patterns identified in the numerical outcomes outlined in section \ref{sec:numerical setup}. It effectively captures the principal observations solely using the initial conditions: (a) The large $j_z$ changes are approximately quantitatively reproduced for a large phase space of $C^0_K$. In particular, the suppression of the effect as $\beta$ becomes larger than $\beta_0$ is recovered. (b) Like in the numerical results $\Delta j_z$ is determined mainly by $C^0_K$. (c) The rich variety of types of dependence of $\Delta j_z$ on $C^0_K$ is mainly generated also by the analytic solution: The distinct division to separate sub-classes and the regions where a significant scatter is observed. Note that there are features that are not reproduced. In particular, a strikingly narrow correlation is observed in the numerical results for $C^0_K>-0.6$ in some of the panels and is not reproduced. This very tight correlation in the numerical results is surprising and warrants further investigation which is beyond the scope of this work. Additionally, in the two lower left panels there are localized increases (at $C^0_K\sim-0.3$) that are not recovered.

\subsection{Distinct Qualitative Pairs of $V$ and $E$\label{subsec:qualitative_V_E}}

The analytic solution assigns each set of parameters and initial conditions to one of four distinct classes defined below. The potential $V$ in Equation \ref{eq:Potential_V} has two distinct shapes depending on whether $C_1$ is smaller or larger than a critical value (using Equation \ref{eq:gamma})
\begin{equation}
 C_{1}^{\text{crit}}=\frac{9}{2}\left(\frac{5}{4}\right)^{\frac{1}{3}}\gamma^{\frac{2}{3}}.
\end{equation}
If $C_{1}<C_{1}^{\text{crit}}$ the potential $V$ has no maxima and one minimum. If $C_{1}>C_{1}^{\text{crit}}$ the potential has a local maximum (and two minima) and we denote by $V_{\text{max}}$ the value of the potential at this local maximum and by $\delta_{V_{\text{max}}}$ the $\delta$ at which $V_{\text{max}}$ is obtained. The properties of Equation \ref{eq:Potential_V} imply that $\delta_{V_{\text{max}}}>0$.

\begin{figure}
 \begin{centering}
  \includegraphics[scale=0.22]{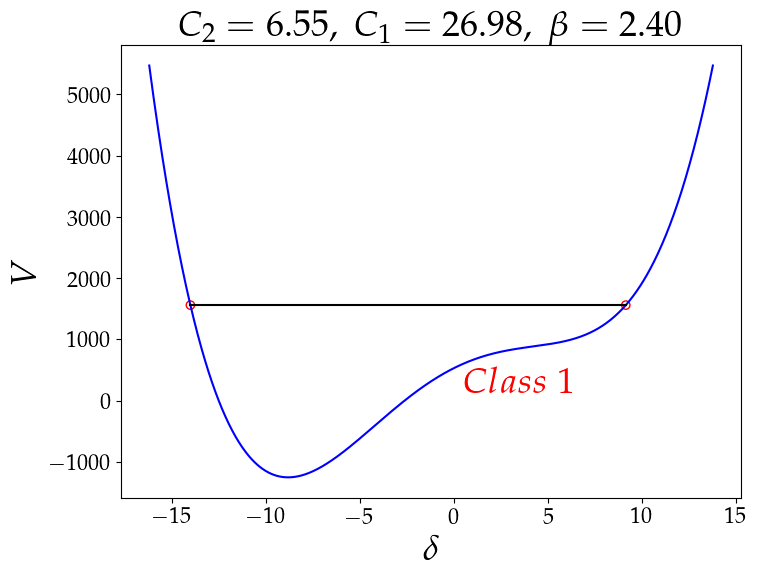}\includegraphics[scale=0.22]{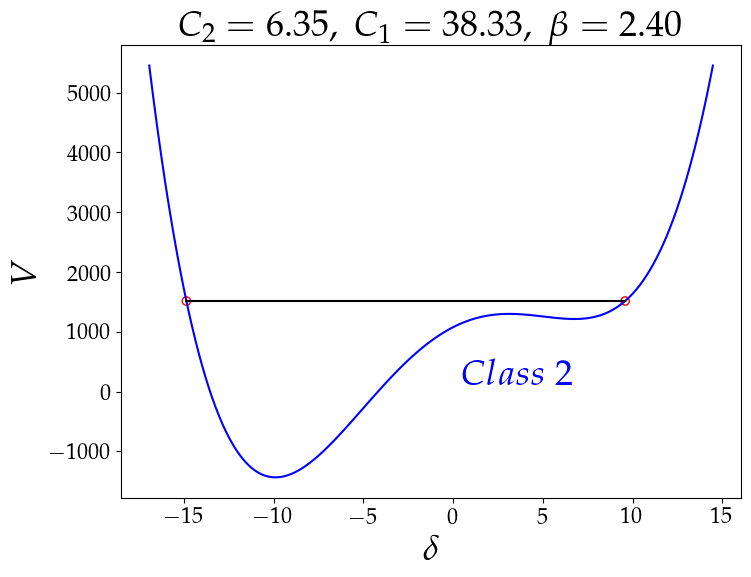}
  \par\end{centering}
 \begin{centering}
  \includegraphics[scale=0.22]{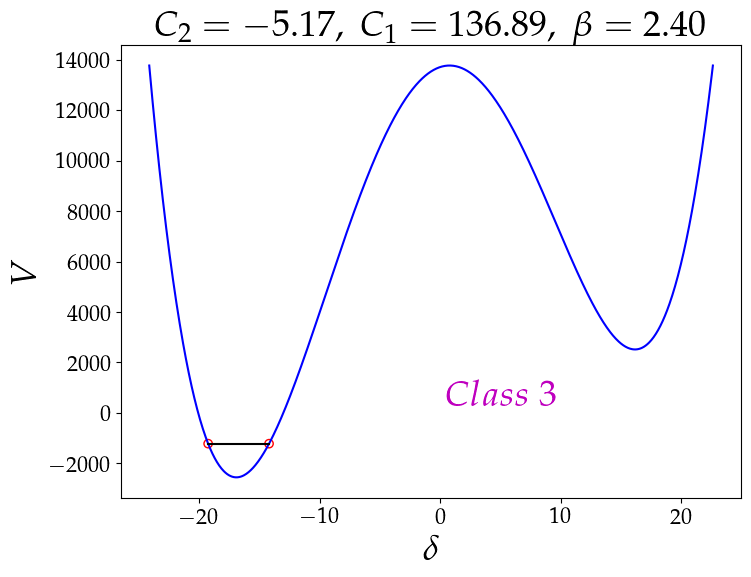}\includegraphics[scale=0.22]{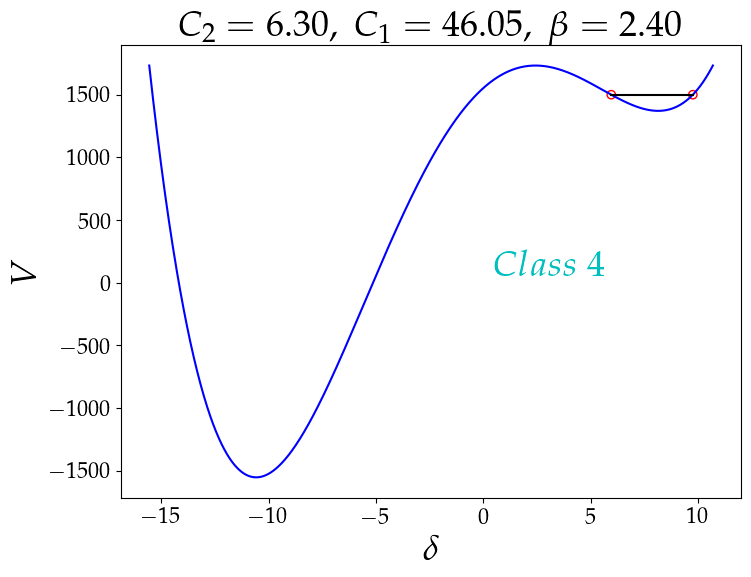}
  \par\end{centering}
\caption{The potential $V$ (Equation \ref{eq:Potential_V}) in blue and the constant energy $E$ (Equation \ref{eq:energy}) in black for $\beta=2.4$ showing the four distinct classes of $\left(V,E\right)$ pairs: upper left panel: Class 1 - no $V_{\text{max}}$, upper right panel: Class 2 - $E>V_{\text{max}}$, lower left panel: Class 3 - $E<V_{\text{max}}$ and $\delta<\delta_{V_{\text{max}}}$ and lower right panel: Class 4 - $E<V_{\text{max}}$ and $\delta>\delta_{V_{\text{max}}}$.\label{fig:potential_energy_cases_beta_2.4}}
\end{figure}

\begin{figure}
 \begin{centering}
 \includegraphics[scale=0.21]{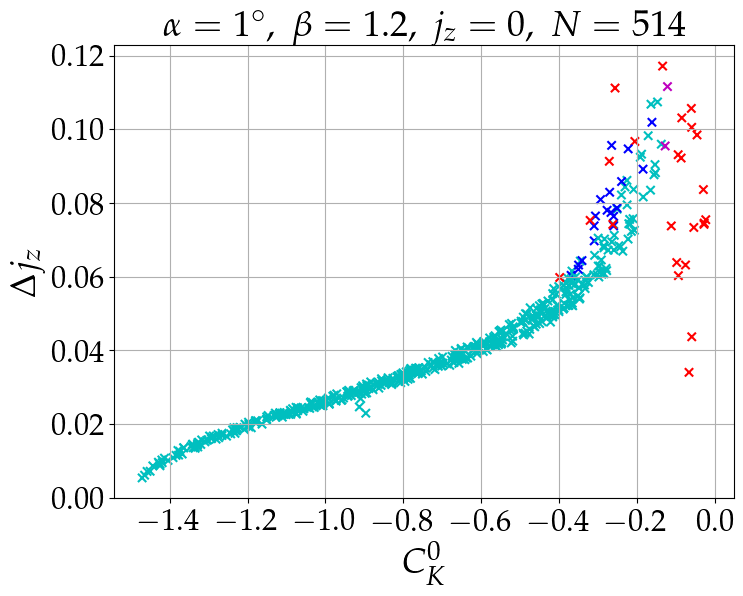}
  \includegraphics[scale=0.21]{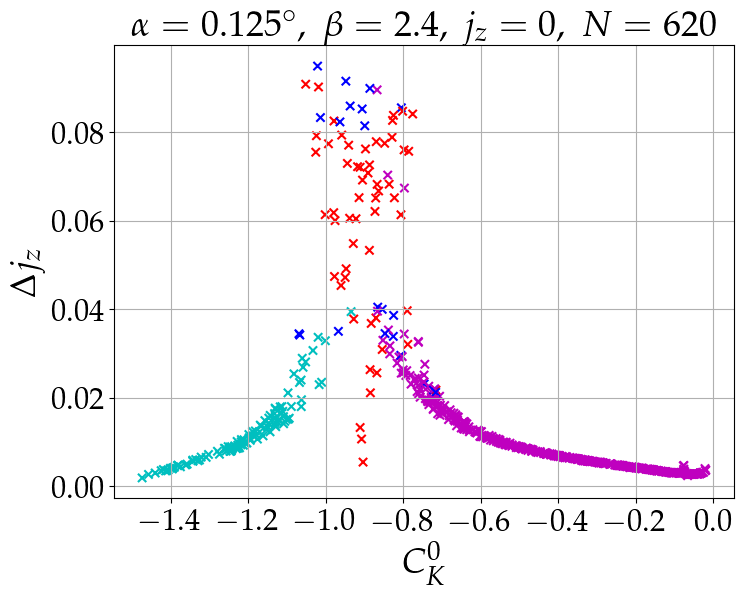}
  \par\end{centering}
 \begin{centering}
 \includegraphics[scale=0.21]{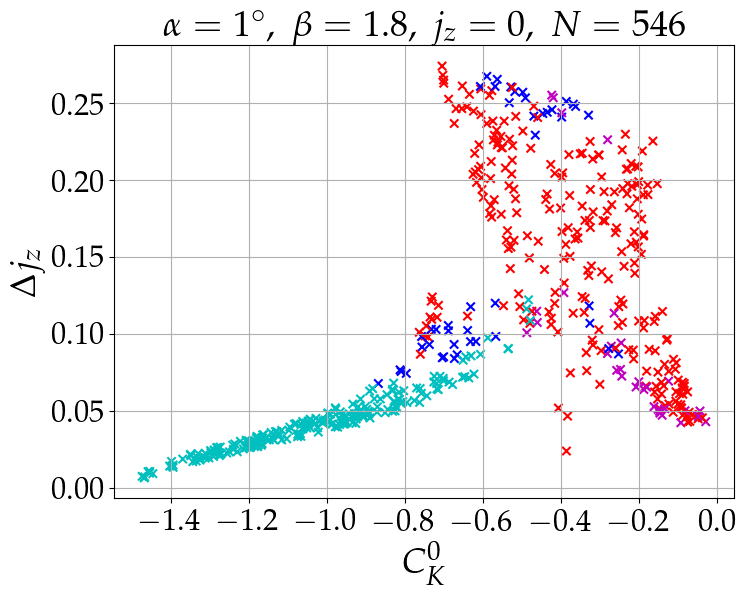}
  \includegraphics[scale=0.21]{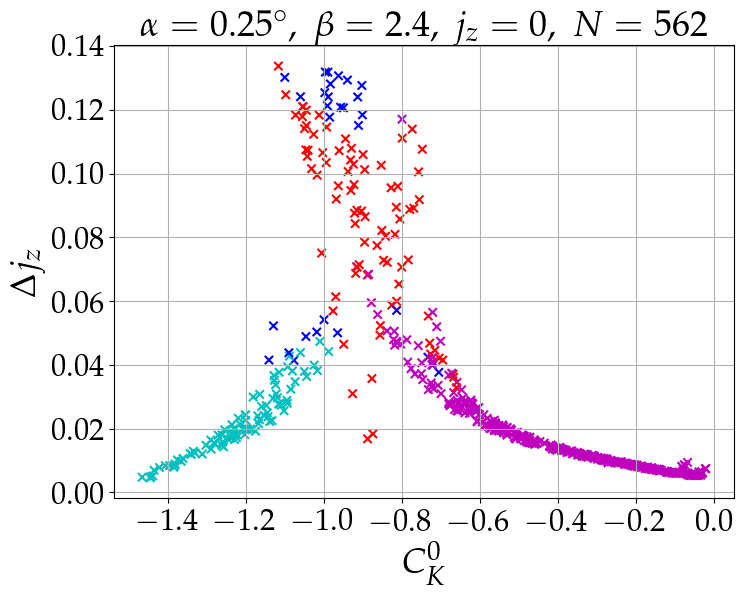}
  \par\end{centering}
   \begin{centering}
 \includegraphics[scale=0.21]{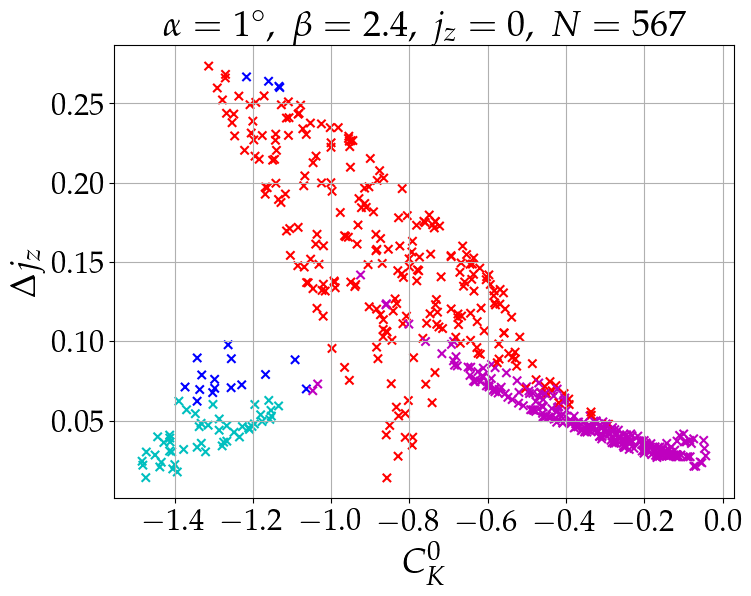}
 \includegraphics[scale=0.21]{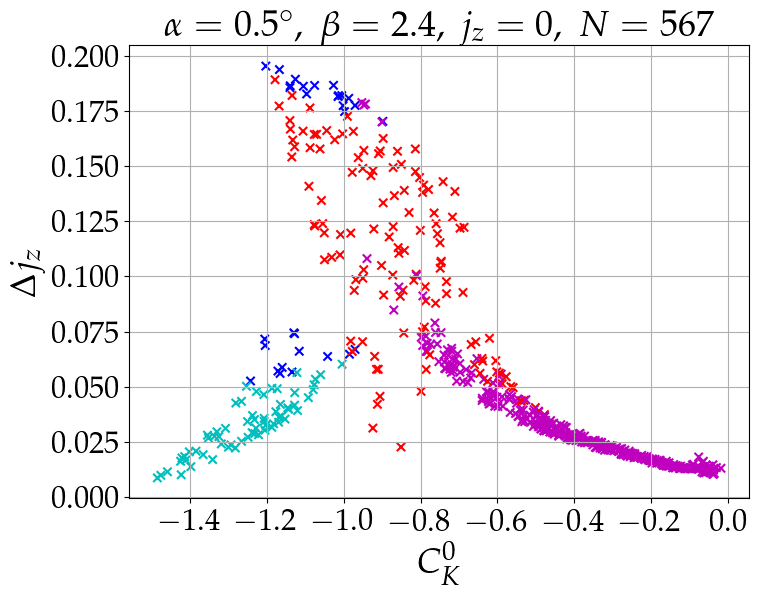}
  \par\end{centering}
   \begin{centering}
   \includegraphics[scale=0.21]{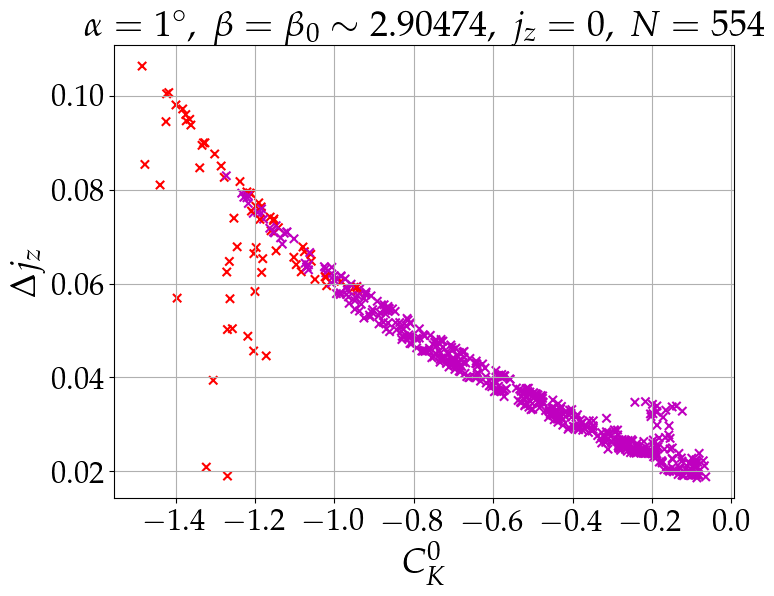}
   \includegraphics[scale=0.21]{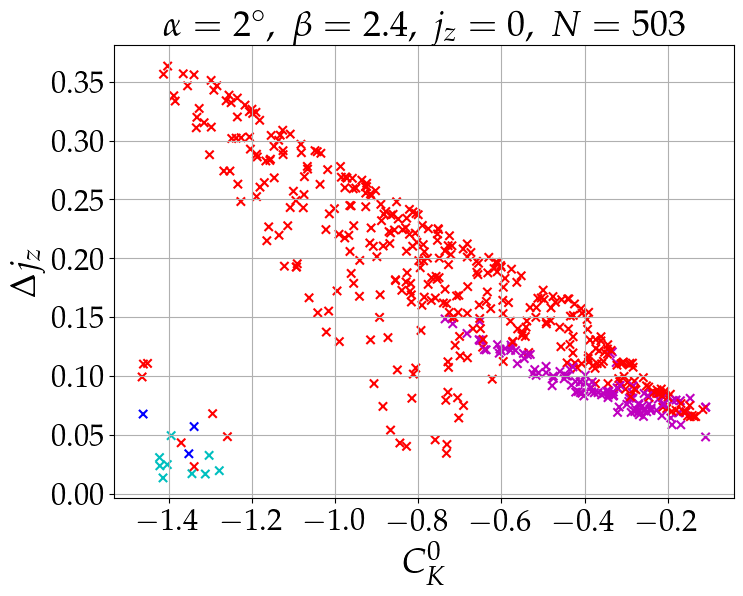}
  \par\end{centering}
     \begin{centering}
   \includegraphics[scale=0.21]{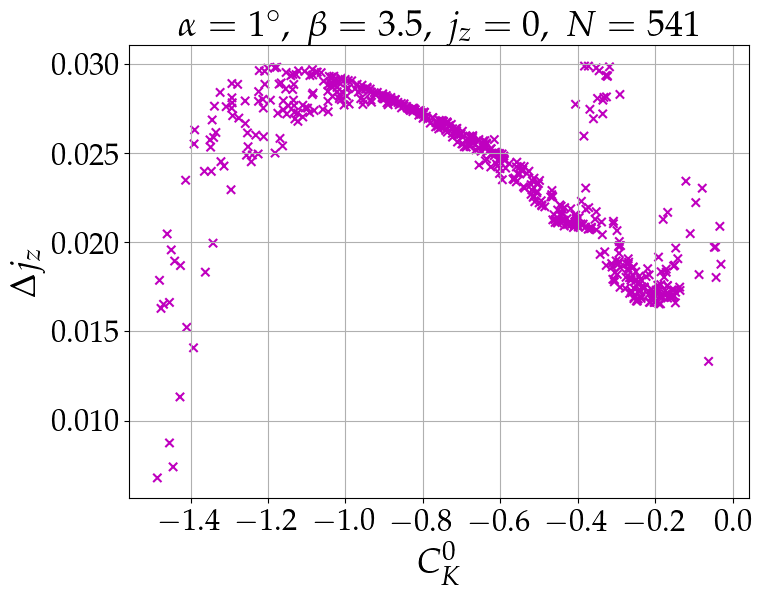}
   \includegraphics[scale=0.21]{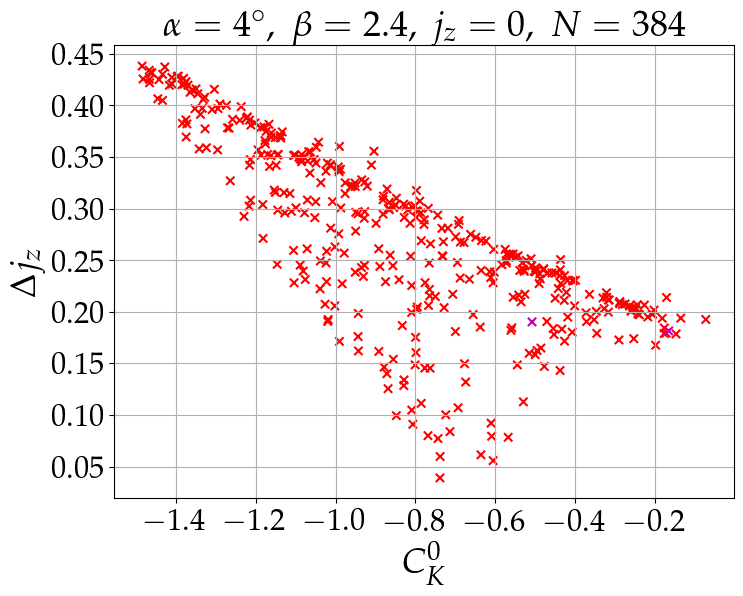}
  \par\end{centering}  
 \caption{Amplitude of the change in $j_z$ vs. initial $C_K$ in the numerical integrations (same as the black crosses in Figure \ref{fig:jz_amplitude_vs_CK_always_librating}) color coded by the analytically specified nature of the pair $\left(V,E\right)$ as defined in section \ref{subsec:qualitative_V_E}: red - Class 1 (no $V_{\text{max}}$), blue - Class 2 ($E>V_{\text{max}}$), magenta - Class 3 ($E<V_{\text{max}}$ and $\delta<\delta_{V_{\text{max}}}$) and cyan - Class 4 ($E<V_{\text{max}}$ and $\delta>\delta_{V_{\text{max}}}$). Panels as in Figure \ref{fig:jz_amplitude_vs_CK_always_librating}.\label{fig:jz_amplitude_vs_CK_always_librating_numerical_only_color_coded}}
\end{figure}

The phase space of initial conditions is separated to four distinct classes (colors correspond to Figures \ref{fig:potential_energy_cases_beta_2.4}, \ref{fig:jz_amplitude_vs_CK_always_librating_numerical_only_color_coded} and \ref{fig:jzamplitude_jzmax_jzmin_vs_Ck_only_librating_colored_by_classes_V_E_pairs_left_numerical_right_analytical}):

\textbf{{\color{red} Class 1} - $C_{1}<C_{1}^{\text{crit}}$ - no $V_{\text{max}}$:} see example in the left upper panel of Figure \ref{fig:potential_energy_cases_beta_2.4}.

\textbf{{\color{blue} Class 2} - $C_{1}>C_{1}^{\text{crit}}, \ E>V_{\text{max}}$}: see example in the right upper panel of Figure \ref{fig:potential_energy_cases_beta_2.4}.

\textbf{{\color{magenta} Class 3} - $C_{1}>C_{1}^{\text{crit}}, \ E<V_{\text{max}}, \ \delta<\delta_{V_{\text{max}}}$}: see example in the left lower panel of Figure \ref{fig:potential_energy_cases_beta_2.4}.

\textbf{{\color{cyan} Class 4} - $C_{1}>C_{1}^{\text{crit}}, \ E<V_{\text{max}}, \ \delta>\delta_{V_{\text{max}}}$}: see example in the right lower panel of Figure \ref{fig:potential_energy_cases_beta_2.4}.


An immediate relation between these classes of initial conditions and $\Delta j_z$ seems to emerge from the examples shown in Figure \ref{fig:potential_energy_cases_beta_2.4}: Large changes in $j_z$ which correspond to large changes in $\delta$ are expected for classes 1 and 2, and not for classes 3 and 4. As shown in section \ref{sec:comparison} this is true for a wide range of initial conditions and parameters.
 
\section{Comparison of trends between analytic and numerical solutions\label{sec:comparison}}

In the following subsections we use the analytic solution to explain the main different structures shown in Figure \ref{fig:jz_amplitude_vs_CK_always_librating}.

The numerical results shown in Figure \ref{fig:jz_amplitude_vs_CK_always_librating} are reproduced in Figure \ref{fig:jz_amplitude_vs_CK_always_librating_numerical_only_color_coded} with different colors corresponding to the analytical classification of the initial conditions defined in section \ref{subsec:qualitative_V_E}.
We note several striking observations: (a) Different numerical features correspond to distinct analytical classes (note a slight exception for Class 2 (blue) in some of the panels which will be discussed in section \ref{subsec:beta_2.4}). (b) As claimed in section \ref{subsec:qualitative_V_E}, large changes in $j_z$ mostly correspond to Class 1 (red) or Class 2 (blue). (c) The subset characterized by low values of $\Delta j_z$ at the negative end of $C^0_K$ predominantly aligns with Class 4 (cyan). (d) The region demonstrating a notably tight correlation appearing as a rightward tail (high $C^0_K$ and low $\Delta j_z$) corresponds to Class 3 (magenta). (e) Initial conditions identified as Class 1 (red) have typically the largest scatter in $\Delta j_z$.

\subsection{$\alpha$ dependence \label{subsec:alpha_dependence}}
The right column of panels in Figures \ref{fig:jz_amplitude_vs_CK_always_librating} and \ref{fig:jz_amplitude_vs_CK_always_librating_numerical_only_color_coded} together with the middle left panel show the results with a constant value of $\beta=2.4$ and a logarithmically increasing $\alpha$ from $0.125^\circ$ to $4^\circ$.

The effect of varying $\alpha$ is three fold:
\begin{itemize}
    \item As $\alpha$ increases - the potential amplitude of $\Delta j_z$ increases. As can be seen in these subplots in Figure \ref{fig:jz_amplitude_vs_CK_always_librating} the analytic solution captures quantitatively the dependence of the amplitude of the effect on $\alpha$. Notably, the \textit{y}-axis scale varies significantly across these subplots, and the green circles track the trends observed in the black crosses.

    \item Increasing $\alpha$ induces changes in the color distributions shown in the maps, with the portion of initial conditions attributed to Class 1 (red) increasing with $\alpha$ and reducing the number of occurrences of classes 3 (magenta) and 4 (cyan). The heightened potential amplitude of $\Delta j_z$ with increasing $\alpha$ can be ascribed to a dual effect: the prevalence of Class 1 cases with a greater amplitude of $\delta$, and the influence of the $\alpha^{\frac{2}{3}}$ term in Equation \ref{eq:C4}, which translates changes in $\delta$ into $\Delta j_z$.

    \item As $\alpha$ increases the count $N$ of librating cases experiences a decline. This phenomenon is rooted in the intensified perturbing potential, causing more vigorous fluctuations in $C_K$ and subsequently leading to an increased occurrence of cases where $C^{\text{max}}_K>0$.
\end{itemize}

\subsection{$\beta$ dependence \label{subsec:beta_dependence}}

The left column of panels in Figures \ref{fig:jz_amplitude_vs_CK_always_librating} and \ref{fig:jz_amplitude_vs_CK_always_librating_numerical_only_color_coded} show the results with a constant value of $\alpha=1^\circ$ and a range of $\beta$ values from $1.2$ to $3.5$ (including $\beta_0\approx2.9$) for which considerable change in $\Delta j_z$ ($\gtrsim0.1$ for $\alpha=1^\circ$) is obtained.

The change of $\beta$ predominantly impacts the distribution of the different classes based on the initial conditions (i.e $C^0_K$). As evidenced by these subplots in Figure \ref{fig:jz_amplitude_vs_CK_always_librating_numerical_only_color_coded}, at $\beta=1.2$ the majority of initial conditions fall into Class 4 (cyan). As $\beta$ increases, the proportion of Class 4 (cyan) gradually diminishes, giving way to an increasing fraction of Class 1 (red) and 3 (magenta). Specifically, for $\beta$ values of $1.8$ and $2.4$, a substantial portion of the phase space is occupied by Class 1 (red), enabling significant changes in $\delta$ and consequently yielding high values of $\Delta j_z$. When $\beta\ge\beta_0$ no instances of Class 2 (blue) or 4 (cyan) are present, as actual initial conditions for $\mathbf{e}$ invariably satisfy $0\le\bar{e}^2_z\le 1$. Consequently, from Equation \ref{eq:delta}, it becomes evident that $\delta<0$ for $\beta\ge\beta_0$, restricting the cases to only classes 1 (red) or 3 (magenta). Further elevating $\beta$ away from $\beta_0$, for instance, to $\beta=3.5$, leads to all cases being categorized as Class 3 (magenta). This phenomenon emerges as a common trait of sufficiently large $\beta$, as an increased minimal value for $\delta^2$ is observed with rising $\beta$ (see Equation \ref{eq:delta}), and consequently, $C_1$ (linked to $\delta^2$, see Equation \ref{eq:C1}) quickly surpasses $C^{\text{crit}}_1$ for all initial conditions, making Class 1 (defined by $C_1<C^{\text{crit}}_1$) unattainable.

\begin{figure}
 \begin{centering}
  \includegraphics[scale=0.4]{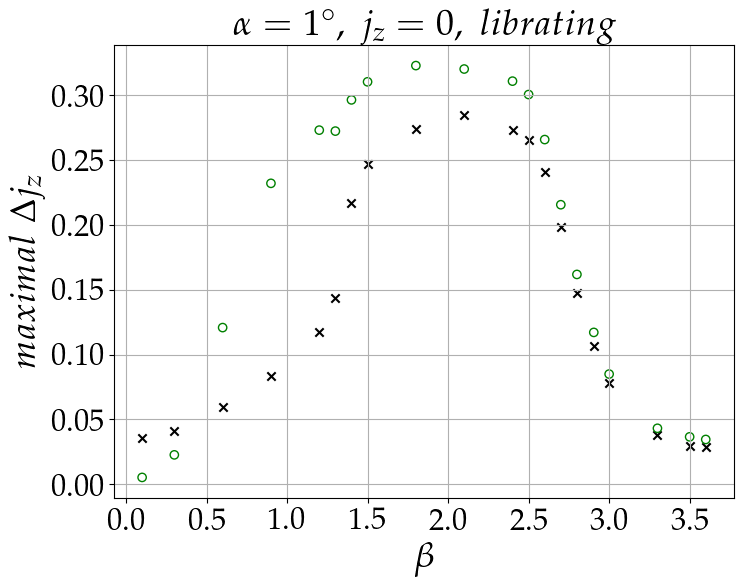}
  \par\end{centering}
 \caption{The maximal amplitude of change in $j_z$ vs. $\beta$ for $\alpha=1^\circ$. Shown are results from full numerical solutions in librating KLCs (black crosses) and as predicted from initial conditions by the analytic solution (green circles). The results for $\beta=1.2,1.8,2.4,\beta_0\approx 2.9$ and $3.5$ are the highest values attained by $\Delta j_z$ in the corresponding panels of Figures \ref{fig:jz_amplitude_vs_CK_always_librating}.\label{fig:maximal_Delta_jz_vs_beta}}
\end{figure}

Generally, these subplots in Figure \ref{fig:jz_amplitude_vs_CK_always_librating} demonstrate that the analytic solution captures quantitatively the dependence of $\Delta j_z$ on $\beta$ as the green circles follow the generic trends of the black crosses (although failing at low values of $\left|C^0_K\right|$ for $\beta=1.2$). The highest values attained by $\Delta j_z$ in these panels is shown as a function of $\beta$ in Figure \ref{fig:maximal_Delta_jz_vs_beta} along with results from additional values of $\beta$ for $\alpha=1^\circ$ and the corresponding analytic results. As in Figure \ref{fig:jz_amplitude_vs_CK_always_librating} black crosses denote numerical results and green circles denote the analytic prediction using the initial conditions. Note the significant elevation of maximal $\Delta j_z$ in the range $1\lesssim\beta\lesssim3$ and the steep decline for $\beta>\beta_0$ as mentioned above. The analytic solution approximately reproduces the numerical result. A large deviation is observed at $\beta\sim1$ and corresponds to the large values of $\Delta j_z$ in the analytic prediction around $C^0_K\sim-0.3$ as can be seen in the upper left panel of Figure \ref{fig:jz_amplitude_vs_CK_always_librating} showing the results for $\beta=1.2$. Note that for lower $C^0_K$ values the agreement between the solutions is much better for this $\beta$.

\subsection{Detailed comparison for a representative case - $\alpha=1^\circ$, $\beta=2.4$\label{subsec:beta_2.4}}

\begin{figure}
 \begin{centering}
 \includegraphics[scale=0.21]{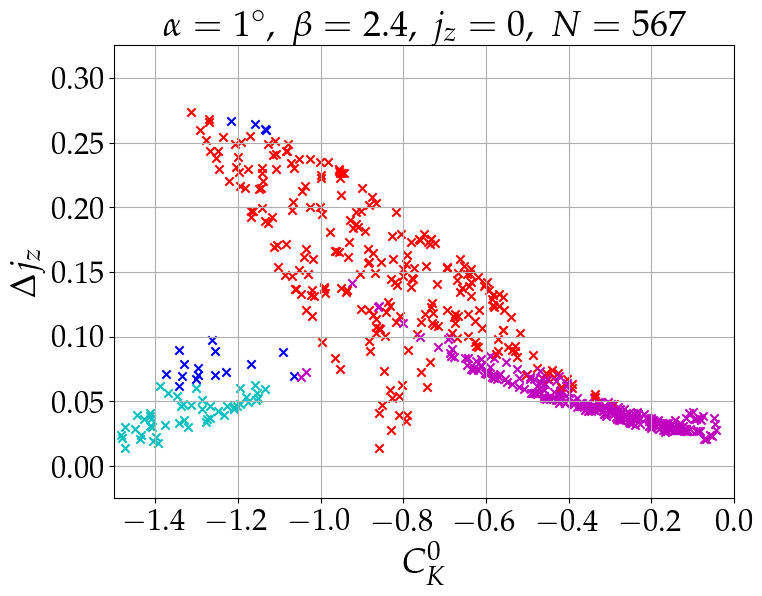}
 \includegraphics[scale=0.21]{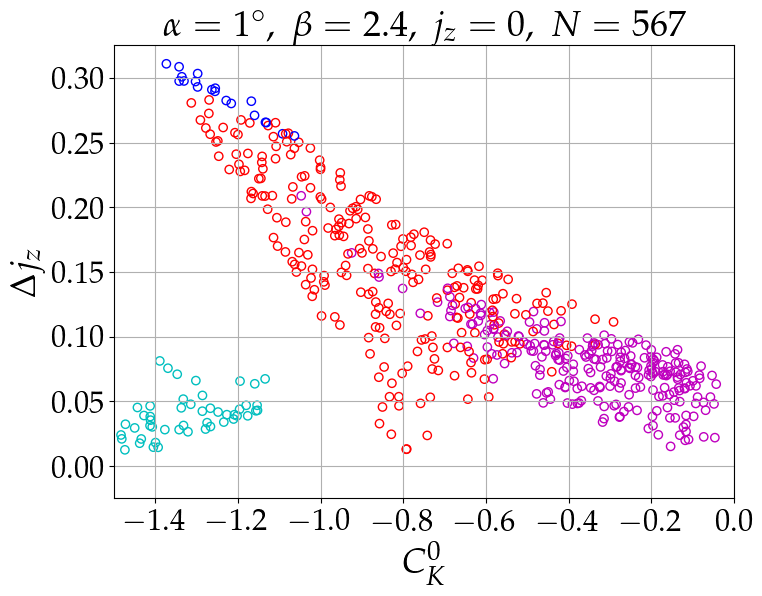}
  \par\end{centering}
 \begin{centering}
  \includegraphics[scale=0.21]{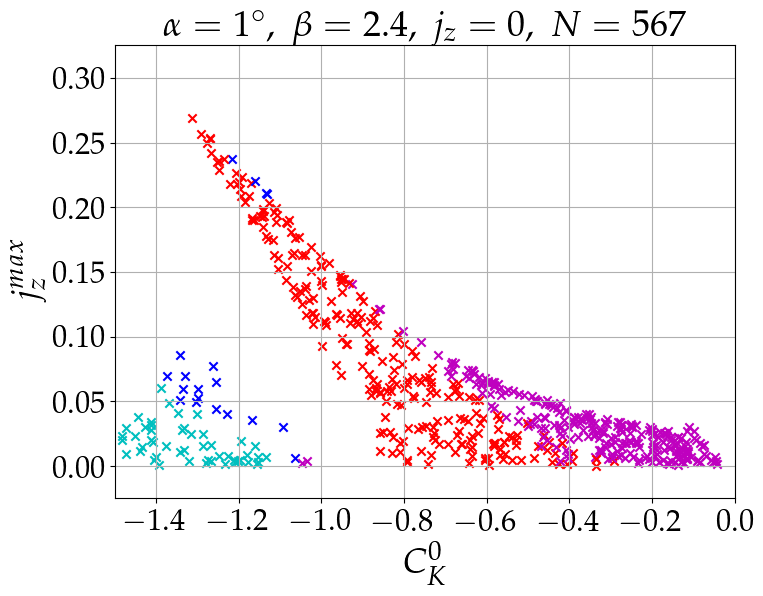}
  \includegraphics[scale=0.21]{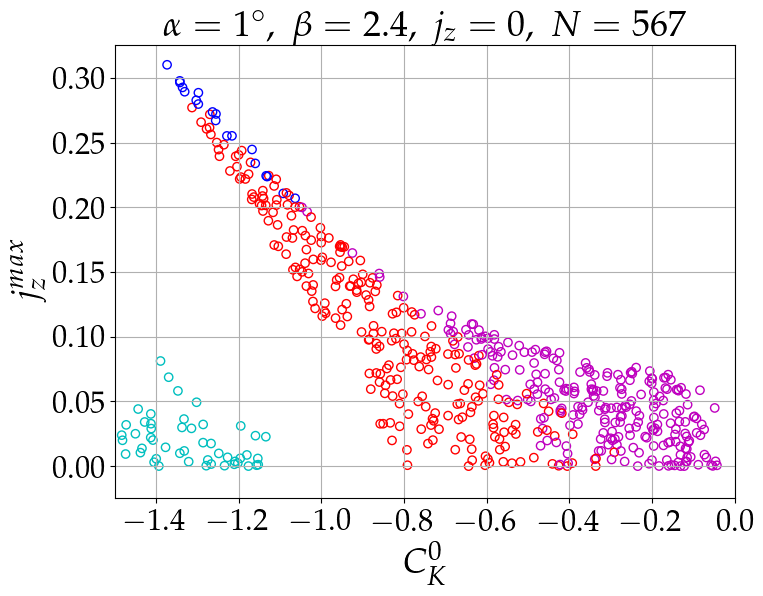}
  \par\end{centering}
  \begin{centering}
  \includegraphics[scale=0.205]{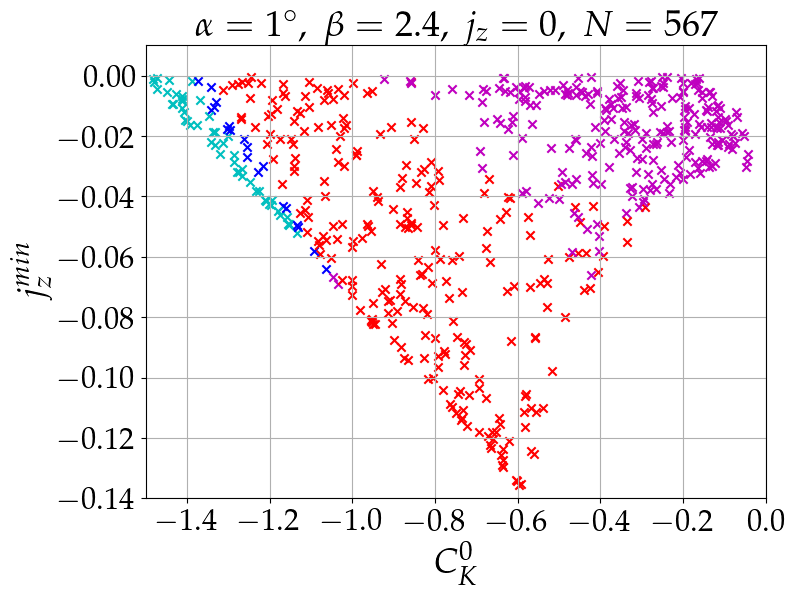}
  \includegraphics[scale=0.205]{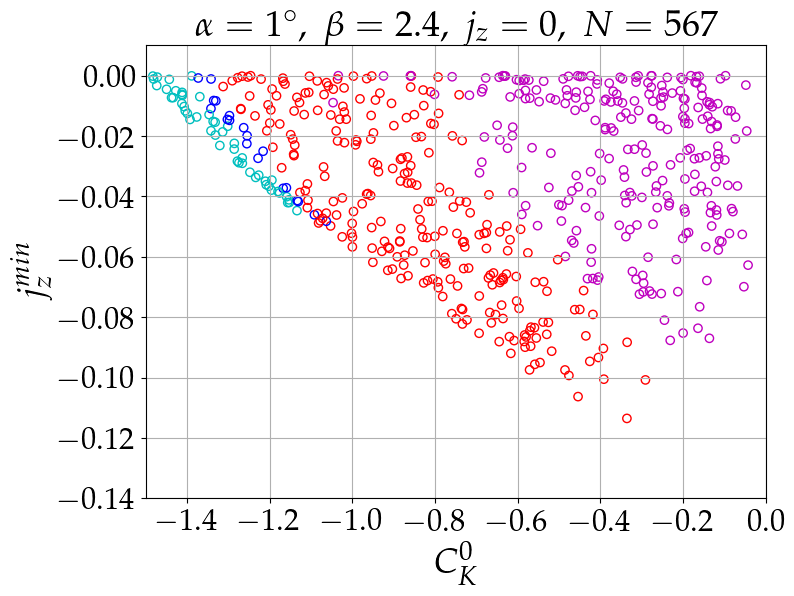}
  \par\end{centering}
 \caption{Results for librating KLCs are shown for a case of $\alpha=1^\circ$ and $\beta=2.4$. Upper panels: Amplitude of the change in $j_z$ vs. initial $C_K$. Middle panels: $j^{\text{max}}_z$ vs. initial $C_K$. Lower panels: $j^{\text{min}}_z$ vs. initial $C_K$. Left column: numerical results. Right column: analytic prediction. The color code in all panels is by the nature of the pair $\left(V,E\right)$ as defined in section \ref{subsec:qualitative_V_E}: red - Class 1 (no $V_{\text{max}}$), blue - Class 2 ($E>V_{\text{max}}$), magenta - Class 3 ($E<V_{\text{max}}$ and $\delta<\delta_{V_{\text{max}}}$) and cyan - Class 4 ($E<V_{\text{max}}$ and $\delta>\delta_{V_{\text{max}}}$). The top left panel is identical to the middle left panel of Figure \ref{fig:jz_amplitude_vs_CK_always_librating_numerical_only_color_coded}.\label{fig:jzamplitude_jzmax_jzmin_vs_Ck_only_librating_colored_by_classes_V_E_pairs_left_numerical_right_analytical}}
\end{figure}

A detailed comparison between the numerical and analytical solutions for a representative case of $\alpha=1^\circ$ and $\beta=2.4$ is given in Figure \ref{fig:jzamplitude_jzmax_jzmin_vs_Ck_only_librating_colored_by_classes_V_E_pairs_left_numerical_right_analytical}. We present the numerical results (left column) and analytical results (right column) with the aforementioned variable of $\Delta j_z$ (top panels), and we add the values composing it: $j^{\text{max}}_z$ (middle panels) and $j^{\text{min}}_z$ (bottom panels). This set of $\alpha$ and $\beta$ values is chosen since high $\Delta j_z$ is reached and rich structures are obtained as can be seen in the middle left panel of Figure \ref{fig:jz_amplitude_vs_CK_always_librating_numerical_only_color_coded}.

As observed in the top right panel, Class 4 cases are a distinct subcategory with a low $\Delta j_z$. As apparent in the middle and bottom panels - this distinction appears in $j^{\text{max}}_z$ and not in $j^{\text{min}}_z$. This can be explained by the analytic solution as follows. First, note that using Equation \ref{eq:C4}, $j^{\text{max}}_z$ ($j^{\text{min}}_z$) corresponds to $\delta_{\text{min}}$ ($\delta_{\text{max}}$). Second, note that when transitioning from Class 4 to Classes 1 or 2, there is a discontinuity in $\delta_{\text{min}}$ but not in $\delta_{\text{max}}$ due to the shape of the potential as can be seen in Figure \ref{fig:potential_energy_cases_beta_2.4}.

A prominent feature in the top panels is the discrepant location of Class 2 (blue): In the right panel, the analytical solution, all blue circles are in the high $\Delta j_z$ regime as expected (see section \ref{subsec:qualitative_V_E}). However, in the left panel, the numerical solution - blue marks appear both in the high $\Delta j_z$ regime and also mixed with Class 4 (cyan crosses) in the low $\Delta j_z$ regime. This discrepancy arises due to the nature of the potential having a local maximum and the high sensitivity regarding whether $E>V_{\text{max}}$ or not. In the analytic solution, Class 2 represents a small region in the initial conditions phase space which is continuously connected to the region occupied by Class 4 but has a discontinuous outcome in $\delta_{\text{min}}$ as mentioned. It is therefore unsurprising to have some misclassification between Class 2 and 4 leading to a notable deviation in $\Delta j_z$. As expected according to the arguments mentioned above the discrepancy appears in $j^{\text{max}}_z$ and not in $j^{\text{min}}_z$ as can be seen in the middle and bottom panels.

Finally, note that the tight correlation observed in the rightward tail in $\Delta j_z$ is much tighter than in $j^{\text{max}}_z$ and $j^{\text{min}}_z$ separately, implying a correlation between them. While the analytical solution does not reproduce the tight correlation in $\Delta j_z$ it is similarly narrower than the corresponding scatters in $j^{\text{max}}_z$ and $j^{\text{min}}_z$ in isolation.

\section{Summary}

In this work, we have generalized the analytic solution developed in \citet{klein2023} for a precession rate $\beta$ in the vicinity of $\beta_0$. Using the analytic solution we obtain predictions for the amplitude of change in $j_z$ based on the initial conditions only, for librating KLCs. These analytic predictions agree approximately with the quantitative results of a full numerical solution of Equations \ref{eq:jOuter_as_a_function_of_tau}-\ref{eq:secular_equations}. The analytic equations are equivalent to a particle moving in a one dimensional potential. We have shown that different features in the rich structures observed in the numerical results correspond to four classes of initial conditions: (i) No local maximum in the potential, (ii) Energy higher than the local maximum, and two classes with energy below the local maximum in one of two local minima (see section \ref{subsec:qualitative_V_E}).

Strictly speaking, the effective equations (based on the ansatz in Equation \ref{eq:exy-ansatz}) and analytic solution were developed close to the fixed point $C_K=-1.5$ and for precession rates around $\beta_0$ (and for $\alpha\ll1$ and $\left|j^0_z\right|\ll1$). In this regime, $\beta_0$ is the natural resonant frequency of the equations. Intriguingly, the generalized solution approximately agrees with the numerical outcomes for the majority of librating KLCs, even when $C^0_K$ approaches values near 0, far from the fixed point and across a wide range of $\beta$ values.

In future work we plan to analytically investigate the rotating KLCs (i.e where $C_K$ stays positive during the evolution). Numerically, the rotating class also shows rich structures and large changes in $j_z$ (not shown here). Given the success of the analysis around the fixed point for librating KLCs, we plan to try a similar approach for rotating KLCs by studying the vicinity of its fixed points at $C_K=1$ (specifically, $j=0$, $e_z=0$, $e_x=\cos\left(\varphi\right)$, $e_y=\sin\left(\varphi\right)$ for any $\varphi$). Finally, there is a regime of initial conditions where $C_K$ changes its sign during the evolution. This class has low $\left|C^0_K\right|$ and has numerically shown a lower amplitude of change in $j_z$ than the other two classes (for $\alpha\sim1^\circ$). For this class other methods should be used for developing an analytic understanding.

\begin{acknowledgments}
We thank Scott Tremaine for a useful discussion suggesting to consider a rotating frame.
\end{acknowledgments}

\bibliography{applicability_of_analytic_solution_precessing_potential_KLC}{}
\bibliographystyle{aasjournal}



\end{document}